\documentclass[10pt,twocolumn,twoside]{IEEEtran}
\usepackage{times}
\usepackage{amsbsy}
\usepackage{amsmath}
\usepackage{latexsym}
\usepackage{amssymb}
\usepackage{times}
\usepackage[ansinew]{inputenc}
\usepackage[dvips ]{graphicx}
\usepackage{epic,eepic}
\input epsf
\title{Set-Membership Adaptive Algorithms based on Time-Varying Error
Bounds for Interference Suppression }
\author{Rodrigo C. de Lamare, IEEE Member and Paulo S. R. Diniz, IEEE Fellow   \\
\thanks{\footnotesize  R. C. de Lamare is with the Communications
Research Group, Department of Electronics, University of York,
United Kingdom and Paulo S. R. Diniz is with LPS/COPPE-UFRJ, Rio
de Janeiro, Brazil. E-mails: rcdl500@ohm.york.ac.uk,
diniz@lps.ufrj.br. } }
\date{  }
\begin{document}
\maketitle
\begin{abstract}
This work presents set-membership adaptive algorithms based on
time-varying error bounds for CDMA interference suppression. We
introduce a modified family of set-membership adaptive algorithms
for parameter estimation with time-varying error bounds. The
algorithms considered include modified versions of the
set-membership normalized least mean squares (SM-NLMS), the affine
projection (SM-AP) and the bounding ellipsoidal adaptive
constrained (BEACON) recursive least-squares technique. The
important issue of error bound specification is addressed in a new
framework that takes into account parameter estimation dependency,
multi-access and inter-symbol interference for DS-CDMA
communications. An algorithm for tracking and estimating the
interference power is proposed and analyzed. This algorithm is
then incorporated into the proposed time-varying error bound
mechanisms. Computer simulations show that the proposed algorithms
are capable of outperforming previously reported techniques with a
significantly lower number of parameter updates and a reduced risk
of overbounding or underbounding.
\\


\end{abstract}

\section{Introduction}

Set-membership filtering (SMF) \cite{diniz,huang,werner1,werner2}
represents a class of recursive estimation algorithms that, on the
basis of a pre-determined error bound, seeks a set of parameters
that yield bounded filter output errors. These algorithms have
been used in a variety of applications such as adaptive
equalization \cite{gollamudi} and multi-access interference
suppression \cite{nagaraj,gollamudi2}. The SMF algorithms are able
to combat conflicting requirements such as fast convergence and
low misadjustment by introducing a modification on the objective
function. In addition, these algorithms exhibit reduced complexity
due to data-selective updates, which involve two steps: a)
information evaluation and b) update of parameter estimates. If
the filter update does not occur frequently and the information
evaluation does not involve much computational complexity, the
overall complexity can be significantly reduced.

The adaptive SMF algorithms usually achieve good convergence and
tracking performance due to an adaptive step size for each update,
and reduced complexity resulting from data selective updating.
However, the performance of SMF techniques depends on the
error-bound specification, which is very difficult to obtain in
practice due to the lack of knowledge of the environment and its
dynamics. In wireless networks characterized by non-stationary
environments, where users often enter and exit the system, it is
very difficult to choose an error bound and the risk of
overbounding (when the error bound is larger than the actual one)
and underbounding (when the error bound is smaller than the actual
one) is significantly increased, leading to performance
degradation. In addition, when the measured noise in the system is
time-varying and the multiple access interference (MAI) and the
intersymbol interference (ISI) encountered by a receiver in a
communication system are highly dynamic, the selection of an
error-bound is further complicated. This is especially relevant
for low-complexity estimation problems encountered in applications
such as mobile terminals and wireless sensor networks
\cite{akyldiz}, where the sensors have limited signal processing
capabilities and power consumption is of central importance. These
problems suggest the deployment of mechanisms to automatically
adjust the error bound in order to guarantee good performance and
a small update rate (UR). It should also be remarked that most of
prior work on adaptive algorithms for interference suppression
\cite{honig} is restricted to systems with short codes. However,
the proposed adaptive techniques are also applicable to systems
with long codes provided some modifications are carried out. For
downlink scenarios, the designer can resort to chip equalization
\cite{klein} followed by a despreader. For an uplink solution,
channel estimation algorithms for aperiodic sequences
\cite{buzzi,xu-long} are required and the sample average approach
for estimating the covariance matrix ${\bf R}=E[{\bf r}(i){\bf
r}^{H}(i)]$ of the observed data ${\bf r}(i)$ has to be replaced
by $\hat{\bf R} = {\bf P}{\bf P}^{H} + \sigma^2{\bf I}$, which is
constructed with a matrix ${\bf P}$ containing the effective
signature sequence of users and the variance $\sigma^{2}$ of the
receiver's noise \cite{liu&xu}.

Previous works on time-varying error bounds include the tuning of
noise bounds in \cite{norton,gazor}, the approach in \cite{deller}
which assumes that the "true" error bound is constant, and the
parameter-dependent error bound recently proposed in
\cite{guo,guo2} with frequency-domain estimation algorithms. The
techniques so far reported do not introduce any mechanism for
tracking the MAI and the ISI and incorporating their power
estimates in the error bound. In addition, the existing approaches
with time-varying bounds have not been considered for more
sophisticated adaptive filtering algorithms such as the affine
projection (AP) and the least-squares (LS) based techniques.

In this work, we propose and analyze a low-complexity framework
for tracking parameter evolution and MAI and ISI power levels,
that relies on simple channel and interference estimation
techniques, and encompasses a family of set-membership algorithms
\cite{huang,werner1,nagaraj,deller2} with time-varying error
bounds. Specifically, we present modified versions of the
set-membership normalized least mean squares (SM-NLMS)
\cite{huang}, the affine projection \cite{werner1} (SM-AP) and the
bounding ellipsoidal adaptive constrained (BEACON)
\cite{nagaraj,deller2} recursive least-squares (RLS) algorithm for
parameter estimation. Then, we incorporate the proposed mechanisms
of interference estimation and tracking into the time-varying
error bounds. In order to evaluate the proposed algorithms, we
consider a DS-CDMA interference suppression application and
adaptive linear multiuser receivers in situations of practical
interest.

This work is organized as follows. Section II briefly describes
the DS-CDMA system model and linear receivers. Section III reviews
the SMF concept with time-varying error bounds and is devoted to
the derivation of adaptive algorithms. Section IV presents the
framework for time-varying error bounds and the proposed
algorithms for channel, interference estimation and tracking.
Section V is dedicated to the analysis of the algorithms for
channel, amplitude, interference estimation and their tracking.
Section VI shows and discusses the simulations results, while
Section VII gives the conclusions.

\section{DS-CDMA System Model and Linear Receivers}

Let us consider the downlink of a symbol synchronous DS-CDMA
system with $K$ users, $N$ chips per symbol and $L_{p}$
propagation paths \cite{honig}. We assume that the delay is a
multiple of the chip rate, the channel is constant during each
symbol interval and the spreading codes are repeated from symbol
to symbol. The received signal $ r(t)$ after filtering by a
chip-pulse matched filter and sampled at chip rate yields the
$M$-dimensional received vector
\begin{equation}
\begin{split}
{\bf r}[i] & = \sum_{k=1}^{K} A_{k}[i]  {b}_{k}[i] {\bf C}_{k}
{\bf h}[i]  +
 {\boldsymbol{\eta}}[i] + {\bf n}[i],
\end{split}
\end{equation}
where $M=N+L_{p}-1$, ${\bf n}[i] = [n_{1}[i]
~\ldots~n_{M}[i]]^{T}$ is the complex Gaussian noise vector with
zero mean and covariance matrix $E[{\bf n}[i]{\bf n}^{H}[i]] =
\sigma^{2}{\bf I}$, where $(\cdot)^{T}$ and $(\cdot)^{H}$ denote
transpose and Hermitian transpose, respectively. The quantity
$E[\cdot]$ stands for expected value, the user $k$ symbol is
$b_k[i]$ and is assumed to be drawn from a general constellation.
The amplitude of user $k$ is $A_{k}[i]$ and
${\boldsymbol{\eta}}[i]$ is the intersymbol interference (ISI) for
user $k$. The $M\times L_{p}$ convolution matrix ${\bf C}_{k}$
that contains one-chip shifted versions of the signature sequence
for user $k$ expressed by ${\bf s}_{k} = [a_{k}(1) \ldots
a_{k}(N)]^{T}$ and the $L_{p}\times 1$ vector ${\bf h}[i]$  with
the multipath components are described by:
\begin{equation}
{\bf C}_{k} = \left[\begin{array}{c c c }
a_{k}(1) &  & {\bf 0} \\
\vdots & \ddots & a_{k}(1)  \\
a_{k}(N) &  & \vdots \\
{\bf 0} & \ddots & a_{k}(N)  \\
 \end{array}\right],
 {\bf h}[i]=\left[\begin{array}{c} {h}_{0}[i]
\\ \vdots \\ {h}_{L_{p}-1}[i]\\  \end{array}\right].
\end{equation}
In this model, the ISI span and contribution ${\boldsymbol
\eta}_k[i]$ are functions of the processing gain $N$ and $L_p$. If
$1< L_p \leq N$ then $3$ symbols would interfere in total, the
current one, the previous and the successive symbols. In the case
of $N < L_p \leq 2N$ then $5$ symbols would interfere, the current
one, the $2$ previous and the $2$ successive ones. In most
practical CDMA systems, we have that $1< L_p \leq N$ and then only
$3$ symbols are usually affected. The reader is referred to UMTS
channel models \cite{umts}, which reveal that the channel usually
affects at most $3$ symbols (it typically spans a few chips).

The multiuser linear receiver design corresponds to determining an
FIR filter ${\bf w}_{k}[i] = \big[w_0[i] ~ w_1 [i]~\ldots~w_{M-1}
\big]^T$ with $M$ coefficients that provides an estimate of the
desired symbol as given by
\begin{equation}
\hat{b}_{k}[i] = \textrm{sgn}\Big(\Re\Big[{\bf w}_{k}^{H}[i]{\bf
 r}[i]\Big]\Big)=  \textrm{sgn}\Big(\Re\Big[{ z}_{k}[i]\Big]\Big),
\end{equation}
where the quantity $\Re(\cdot)$ selects the real part and
$\textrm{sgn}(\cdot)$ is the signum function. The quantity
$z_{k}[i] = {\bf w}_{k}^{H}[i]{\bf r}[i]$ is the output of the
receiver parameter vector ${\bf w}_{k}$ for user $k$, which is
optimized according to a chosen criterion.

\section{Set-Membership Adaptive Algorithms with Time-Varying Error Bounds and Problem Statement}

In this section, we describe a framework that encompasses modified
set-membership (SM) adaptive algorithms with time-varying error
bounds for communications applications. In an SM filtering
\cite{huang} framework, the parameter vector ${\bf w}_{k}[i]$ for
user $k$ in a multi-access system is designed to achieve a
specified bound on the magnitude of the estimation error
$e_{k}[i]=b_{k}[i] - {\bf w}_{k}^{H}[i]{\bf r}[i] $. As a result
of this constraint, the SM adaptive algorithm will only perform
filter updates for certain data. Let $\Theta_k [i]$ represent the
set containing all ${\bf w}_{k}[i]$ that yields an estimation
error upper bounded in magnitude by a time-varying error bound
$\gamma_k[i]$. Thus, we can write
\begin{equation}
\Theta_k[i] = \bigcap_{{(b_k[i], ~{\bf r}[i])\in {\boldsymbol
S}_k}} \{ {\bf w}_{k} \in {\mathcal{C}}^{M}:\mid e_{k}[i]\mid \leq
\gamma_k[i] \},
\end{equation}
where ${\bf r}[i]$ is the observation vector, ${\boldsymbol S}_k$
is the set of all possible data pairs $(b_{k}[i],~ {\bf r}[i])$
and the set $\Theta_k[i]$ is referred to as the feasibility set
for user $k$, and any point in it is a valid estimate
${z}_{k}[i]={\bf w}_{k}^{H}[i]{\bf r}[i]$. Since it is not
practical to predict all data pairs, adaptive methods work with
the membership sets $\psi_{k,i}= \bigcap_{m=1}^{i}
{\mathcal{H}}_{k,m}$ provided by the observations, where
${\mathcal{H}}_{k,m}=\{{\bf w}_{k} \in {\mathcal{C}}^{M}:
|b_{k}[m] - z_{k}[m]| \leq \gamma_k[m]\}$ is the constraint set.
It can be seen that the feasibility set $\Theta_k[i]$ is a subset
of the exact membership set at any given time instant. The
feasibility set $\Theta_k[i]$ is also the limiting set of the
exact membership set, i.e., the two sets will be equal if the
training signal traverses all signal pairs belonging to
${\boldsymbol S}_k$. The idea of the SM algorithms is to
adaptively find an estimate that belongs to the feasibility set
$\Theta_k[i]$. One alternative is to apply one of the many OBE
algorithms such as the bounding ellipsoidal adaptive constrained
(BEACON) \cite{deller2,nagaraj} recursive least-squares (RLS)
algorithm, which tries to approximate the exact membership set
with ellipsoids. Another way is to compute a point estimate
through projections using, for example, the information provided
by the constraint set ${\mathcal{H}}_{k,i}$, as done by the
set-membership NLMS (SM-NLMS) and the affine projection
\cite{werner1} (SM-AP) algorithms. In order to devise an effective
SM algorithm, the error bound $\gamma_k[i]$ must be appropriately
chosen. Due to time-varying nature of many practical environments,
this error bound should also be adaptive and adjustable to certain
characteristics of the environment for the SM estimation
technique. The natural question that arises is: how to design an
efficient and effective mechanism to adjust $\gamma_k[i]$? In what
follows, we will present a modified family of SM adaptive
algorithms that rely on general time-varying error bounds.
Specifically, we will consider the SM-NLMS \cite{huang}, SM-AP
\cite{werner1} and BEACON \cite{deller2,nagaraj} algorithms and we
will modify them such that they will operate with general
time-varying error bounds.

\subsection{SM-NLMS Algorithm with Time-Varying Bounds}

In order to derive an SM-NLMS adaptive algorithm with time-varying
bounds using point estimates, we consider the following
optimization problem
\begin{equation}
\begin{split}
{\textrm{minimize}} & ~ || {\bf w}_k[i+1] - {\bf w}_k[i]||^2  \\
{\textrm {subject to}} & ~ (b_k[i] - {\bf w}_k^H[i+1]{\bf r}[i]) =
{g}_k[i]
\end{split}
\end{equation}
In order to solve the above constrained optimization problem, we
resort to the method of Lagrange multipliers \cite{diniz,bert},
which yields the unconstrained cost function
\begin{equation}
\begin{split}
{\mathcal{L}} & = || {\bf w}_{k}[i+1]-{\bf w}_{k}[i] ||^{2} + 2\Re
\Big[{\bf \lambda}^{*}(b_{k}[i] - {\bf w}_{k}^{H}[i+1]{\bf
r}[i]-g_{k}[i]) \Big],
\end{split}
\end{equation}
where $*$ denotes complex conjugate, ${\bf \lambda}$ is a Lagrange
multiplier and $g_{k}[i]$ is the time-varying set-membership
constraint for user $k$. Taking the gradient terms of (6) with
respect to ${\bf w}_{k}[i+1]$ and ${\bf \lambda}*$, and setting
them to zero, leads us to a system of equations. Solving these
equations yields:
\begin{equation}
e_{k}[i] = b_{k}[i] - {\bf w}_{k}^{H}[i]{\bf r}[i],
\end{equation}
\begin{equation}
{\bf w}_{k}[i+1] = {\bf w}_{k}[i] + ({\bf r}^{H}[i]{\bf
r}[i])^{-1} (e_{k}[i]-g_{k}[i])^{*}{\bf r}[i],
\end{equation}
where $e_{k}[i]$ is the error for user $k$. By  choosing
$g_{k}[i]$ such that $e_{k}[i]$ lies on the closest boundary of
$\Theta_k[i]$ and considering a time-varying error bound
$\gamma_k[i]$, i.e., $g_{k}[i]=\gamma_k[i] sgn(e_{k}[i])$
\cite{huang}, we obtain the following data dependent update rule
and step size
\begin{equation}
{\bf w}_{k}[i+1] = {\bf w}_{k}[i] + \mu_{w}[i] e_{k}^{*}[i]{\bf
r}[i],
\end{equation}
\begin{equation}
\mu_{w}[i] = \left\{ \begin{array}{ll}
\frac{1}{{\bf r}^{H}[i]{\bf r}[i]} (1 - \gamma_k[i]/|e_{k}^*[i]|) & \textrm{if $|e_{k}^{*}[i]|>\gamma_k[i]$,}\\
0 & \textrm{otherwise.}\\
\end{array}\right.
\end{equation}

\subsection{SM-AP Algorithm with Time-Varying Bounds}

In order to describe a modified  SM-AP algorithm with time-varying
bounds, let us first define the observation matrix ${\bf Y}[i] =
[{\bf r}[i] ~\ldots~ {\bf r}[i-P+1]]$, the desired output vector
${\bf b}_{k}[i] = [b_{k}[i]~\ldots ~b_{k}[i-P+1]]^{T}$ that
comprises $P$ outputs and the error vector
\begin{equation}
{\bf e}_{k}[i] = \left[\begin{array}{c}
{b^{*}_{k}[i]-{\bf r}^{H}[i] {\bf w}_{k}[i]}\\
\vdots \\
{b^{*}_{k}[i]-{\bf r}^{H}[i-P+1]{\bf w}_{k}[i]  } \\
 \end{array}\right] =  {\bf b}^{*}_{k}[i]-{\bf Y}^{H}[i]{\bf
 w}_{k}[i].
\end{equation}
The SM-AP adaptive algorithm with time-varying bounds can be
derived from the optimization problem
\begin{equation}
\begin{split}
{\textrm{minimize}} & ~ || {\bf w}_k[i+1] - {\bf w}_k[i]||^2  \\
{\textrm {subject to}} & ~ ({\bf b}_k[i] - {\bf Y}^H[i]{\bf
w}_k[i+1]) = {\bf g}_k[i].
\end{split}
\end{equation}
In order to solve the above problem, we employ the method of
Lagrange multipliers and consider the unconstrained cost function
\begin{equation}
\begin{split}
{\mathcal{L}} & = ||{\bf w}_{k}[i+1]-{\bf w}_{k}[i]||^{2} +
2\Re\Big[ ({\bf b}_{k}[i] - {\bf Y}^{H}[i]{\bf w}_{k}[i+1] - {\bf
g}_{k}[i])^{H} \boldsymbol{ \lambda}  \Big],
\end{split}
\end{equation}
where $\boldsymbol{ \lambda}$ is the vector with Lagrange
multipliers and ${\bf g}_k[i]$ is a constraint vector. By
calculating the gradient terms of (13) with respect to ${\bf
w}_{k}[i+1]$ and ${\boldsymbol \lambda}$, setting them to zero and
solving the resulting equations we arrive at the following
algorithm:
\begin{equation}
{\bf t}_{k}[i] = ({\bf Y}^{H}[i]{\bf Y}[i] + \delta {\bf I})^{-1}
({\bf e}_{k}[i]-{\bf g}[i]),
\end{equation}
\begin{equation}
{\bf w}_{k}[i+1] = {\bf w}_{k}[i] +  {\bf Y}[i] {\bf t}_{k}[i],
\end{equation}
where $\delta$ is a small constant inserted in addition to the
term ${\bf Y}^{H}[i]{\bf Y}[i]$ for improving robustness. If we
select ${\bf e}_{k}[i] - {\bf g}_{k}[i] = (e_{k}[i] - \gamma
sgn(e_{k}[i])){\bf u} = (1 - \gamma_k[i]/|e_{k}[i]|) e_{k}[i]{\bf
u}$, where the {\it a posteriori} errors $e_{k}[i-j]$ are kept
constant for $j=1,~\ldots,P-1$ and ${\bf u}=[1~ 0~\ldots~0]^{T}$,
we obtain the following recursion for the update of ${\bf
t}_{k}[i]$:
\begin{equation}
{\bf t}_{k}[i] = ({\bf Y}^{H}[i]{\bf Y}[i] + \delta {\bf I})^{-1}
(1 - \gamma/|e_{k}[i]|) e_{k}[i]{\bf u}.
\end{equation}
Substituting (16) into (15) and using the bound constraint, we
obtain the following SM-AP algorithm:
\begin{equation}
{\bf w}_{k}[i+1] = {\bf w}_{k}[i] + \mu_{w}[i] {\bf Y}[i] ({\bf
Y}^{H}[i]{\bf Y}[i] + \delta {\bf I})^{-1} e_{k}[i]{\bf u},
\end{equation}
\begin{equation}
\mu_{w}[i] = \left\{ \begin{array}{ll}
(1 - \gamma_k[i]/|e_{k}[i]|) & \textrm{if $|e_{k}[i]|>\gamma_k[i]$,}\\
0 & \textrm{otherwise.}\\
\end{array}\right.
\end{equation}
The SM-AP algorithm described here has computational complexity of
$UR \times {\mathcal{O}}(PM + 2K_{inv}P^{2} )$, where $K_{inv}$ is
a factor required to invert a $P\times P$ matrix \cite{diniz} and
$UR$ is the update rate. Note that the SM-AP is a generalized case
of the SM-NLMS where $P$ data vectors are used to increase the
convergence speed.

\subsection{BEACON Adaptive Algorithm with Time-Varying Bounds}

Here, we propose a modification for a computationally efficient
version of an optimal bounding ellipsoidal (OBE) algorithm called
the Bounding Ellipsoidal Adaptive Constrained Least-Squares
(BEACON) algorithm \cite{nagaraj}, which is closely related to a
constrained least-squares optimization problem. The proposed
technique amounts to deriving the BEACON algorithm equipped with
time-varying bounds. Unlike the other previously reported
low-complexity algorithms \cite{huang,werner1} and the modified
SM-NLMS and SM-AP techniques described in the previous
subsections, the modified BEACON recursion has the potential to
achieve a very fast convergence performance, which is relatively
independent from the eigenvalue spread of the data covariance
matrix as compared to the stochastic gradient algorithms
\cite{diniz}.

The proposed BEACON algorithm with a time-varying bound can be
derived from the following optimization problem \cite{nagaraj}
\begin{equation}
\begin{split}
{\textrm{minimize}} & ~ \sum_{l=1}^{i-1} \lambda^{i-l}[l] |b_k[l] - {\bf w}_k^H[i]{\bf r}[l]|^2  \\
{\textrm {subject to}} & ~ |{\bf b}_k[i] - {\bf w}_k[i]^H {\bf
r}[i] |^2 = \gamma_k^2[i].
\end{split}
\end{equation}

The constrained problem above can be recast as an unconstrained
one and be solved via an unconstrained least-squares cost function
using the method of Lagrange multipliers given by
\begin{equation}
\begin{split}
{\mathcal{L}} & = \sum_{l=1}^{i-1} \lambda_k[l]^{i-l} |b_k[l] -
{\bf w}^{H}_k[i]{\bf r}[l]|^{2} +  \lambda_k[i](|b_k[i] - {\bf
w}^{H}_k[i]{\bf r}[i]|^2 - \gamma_k^2[i]),
\end{split}
\end{equation}
where $\lambda_k[l]$ plays the role of Lagrange multiplier and
forgetting factor at the same time for user $k$. The solution to
the above optimization problem is obtained by taking the gradient
terms with respect to ${\bf w}_k[i]$ and making them equal to
zero. After some mathematical manipulations we have
\begin{equation}
{\bf w}_k[i] = {\bf w}_k[i-1] + \lambda_k[i] {\bf P}_k[i] {\bf
r}[i] (b_k[i] - {\bf w}^{H}_k[i]{\bf r}[i])^{*}.
\end{equation}

By using the constraint $b_k[i] - {\bf w}^{H}_k[i]{\bf r}[i] = (
\xi_k[i] / |\xi_k[i]| ) \gamma_k[i]= \xi_k[i] - (\lambda_k[i] {\bf
G}_k[i] ( \xi_k[i] / |\xi_k[i]| ) \gamma_k[i])$, and the matrix
inversion lemma \cite{diniz}, we can arrive at the BEACON
algorithm with time-varying bounds described by
\begin{equation}
{\bf P}_k[i] = {\bf P}_k[i-1] - \frac{\lambda_k[i] {\bf P}_k[i-1]
{\bf r}[i]{\bf r}^{H}[i]{\bf P}_k[i-1]}{1+ \lambda_k[i] t_k[i]},
\end{equation}
\begin{equation}
{\bf w}_k[i] = {\bf w}_k[i-1] + \lambda_k[i] {\bf P}_k[i] {\bf
r}[i] \xi^{*}_k[i],
\end{equation}
where the prediction error is $\xi_k[i] = b_k[i] - {\bf
w}^{H}_k[i-1] {\bf r}[i]$, $t_k[i] = {\bf r}^{H}[i]{\bf
P}_k[i-1]{\bf r}[i]$, and $\lambda_k[i] \geq 0$. In order to
compute the optimal value for $\lambda_k[i]$, the algorithm
considers the following cost function \cite{nagaraj}:
\begin{equation}
{\mathcal{C}}_{\lambda_k[i] } = \lambda_k[i] \Bigg[
\frac{\xi_k[i]}{\gamma^2_k[i]} \Bigg( \frac{1}{1+
 \lambda_k[i]t_k[i]} \Bigg) -1 \Bigg].
\end{equation}
The minimization of the cost function in (24) leads to the
innovation check of the proposed BEACON algorithm:
\begin{equation}
\lambda_k[i] = \left\{ \begin{array}{ll}
\frac{1}{t_k[i]} \Bigg( \frac{|\xi_k[i]|}{\gamma_k[i]} - 1 \Bigg) & \textrm{if $|\xi^{*}_k[i]|>\gamma_k[i],$}\\
0 & \textrm{otherwise.}\\
\end{array}\right.
\end{equation}

\section{Algorithms for Time-Varying Error Bounds, Interference Estimation and Tracking}

This section is devoted to time-varying error bounds that
incorporate parameter and interference dependency. We propose a
low-complexity framework for time-varying error bounds,
interference estimation and tracking. A simple and effective
algorithm for taking into consideration parameter dependency is
introduced and incorporated into the error bound. A procedure for
estimating MAI and ISI power levels is also presented and employed
in the adaptive error bound for SM algorithms. The proposed
algorithms are based on the use of simple rules and parameters
that behave as forgetting factors, regulate the pace of time
averages and selectively weigh some quantities.

\subsection{Parameter Dependent Bound}

Here, we describe a parameter dependent bound (PDB), that is
similar to the one proposed in \cite{guo} and considers the
evolution of the parameter vector ${\bf w}_{k}[i]$ for the desired
user (user $k$). The proposed PDB recursion computes a bound for
SM adaptive algorithms and is described by:
\begin{equation} \gamma_k[i+1] = (1-\beta)  \gamma_k[i] +
\beta \sqrt{\alpha||{\bf w}_{k}[i]||^{2}
{\hat{\sigma}}^{2}_{v}[i]},
\end{equation}
where $\beta$ is a forgetting factor that should be adjusted to
ensure an appropriate time-averaged estimate of the evolutions of
the parameter vector ${\bf w}_k[i]$, $\alpha||{\bf w}_{k}[i]||^{2}
{\hat{\sigma}}^{2}_{v}[i]$ is the variance of the inner product of
${\bf w}_{k}[i]$ with ${\bf n}[i]$ that provides information on
the evolution of ${\bf w}_{k}[i]$, $\alpha$ is a tuning parameter
and ${\hat{\sigma}}^{2}_{v}[i]$ is an estimate of the noise power.
This kind of recursion helps avoiding too high or low values of
the squared norm of ${\bf w}_k[i]$ and provides a smoother
evolution of its trajectory for use in the time-varying bound.
The noise power at the receiver should be estimated via a time
average recursion. In this work, we will assume that it is known
at the receiver.

\subsection{Parameter and Interference Dependent Bound}

In this part, we develop an interference estimation and tracking
procedure to be combined with a parameter dependent bound and
incorporated into a time-varying error bound for SM recursions.
The MAI and ISI power estimation scheme, outlined in Fig. 2,
employs both the RAKE receiver and the linear receiver described
in (3) for subtracting the desired user signal from ${\bf r}[i]$
and estimating MAI and ISI power levels. With the aid of adaptive
algorithms, we design the linear receiver, estimate the channel
modeled as an FIR filter for the RAKE receiver and obtain the
detected symbol $\hat{ b}_{k}[i]$, which is combined with an
amplitude estimate $\hat{A}_{k}[i]$ for subtracting the desired
signal from the output $x_{k}[i]$ of the RAKE. Then, the
difference $d_{k}[i]$ between the desired signal and $x_{k}[i]$ is
used to estimate MAI and ISI power.

\begin{figure}[!htb]
\begin{center}
\hspace*{-3.5em}{\includegraphics[width=11.5cm,
height=6cm]{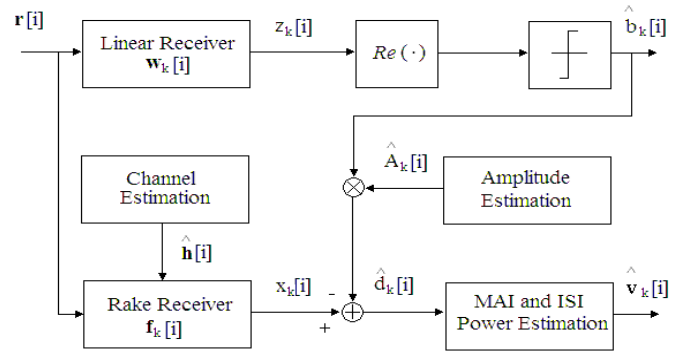}} \vspace*{0.0em} \caption{ Block diagram of
the proposed interference estimation and tracking algorithm.}
\end{center}
\end{figure}

~

\textit{Channel Estimation:} \\

Let us first present a simple channel estimation algorithm for
designing the RAKE receiver. Consider the constraint matrix ${\bf
C}_{k}$ that contains one-chip shifted versions of the signature
sequence for user $k$ defined in (2) and the assumption that the
symbols $b_k[i]$ are independent and identically distributed
(i.i.d) and statistically independent from the symbols of the
other users. The proposed channel estimation algorithm is based on
the following cost function
\begin{equation}
\begin{split}
{\mathcal{C}}(\hat{\bf h}[i],\hat{A}_k[i]) & = E \big[ ||
\hat{A}_k[i] b_k[i]{\bf C}_{k}\hat{\bf h}[i] - {\bf r}[i] ||^2
\big] \\ & = E \big[ || \hat{A}_k[i] b_k[i]\hat{\bf f}_k[i] - {\bf
r}[i] ||^2 \big],
\end{split}
\end{equation}
where $\hat{\bf h}[i]$ is an estimate of the channel ${\bf h}[i]$
and $\hat{\bf f}_k[i] = {\bf C}_{k}\hat{\bf h}[i]$ is the RAKE
receiver for user $k$ with the estimated channel. By taking the
gradient terms of (27), making them equal to zero, we can devise a
stochastic gradient (SG) channel estimation algorithm as follows:
\begin{equation}
\hat{\bf h}[i+1] = \hat{\bf h}[i] - \mu_h \hat{A}_k[i]{\bf C}_k^H
b_k^*[i] \big( b_k[i] {\bf C}_k \hat{\bf h}[i] - {\bf r}[i] \big).
\end{equation}

\textit{Amplitude Estimation:}\\

The amplitude has also to be estimated at the receiver in order to
provide this information for different tasks such as interference
cancellation and power control. The proposed interference
estimation and tracking algorithm needs some form of amplitude
estimation in order to proceed with the estimation of the
interference power. To estimate the amplitudes of the associated
user signals, we describe the following optimization problem with
the cost function in (27)
\begin{equation}
\begin{split}
{\hat A}_{k}[i] & =  \arg \min_{{A}_{k}[i]} {\mathcal{C}}(\hat{\bf
h}[i],\hat{A}_k[i]) \\ & =  \arg \min_{{A}_{k}[i]}
E[||{A}_{k}[i]\hat{b}_k[i] {\bf C}_{k}\hat{\bf h}[i] - {\bf
r}[i]||^{2}].
\end{split}
\end{equation}
In order to solve the problem above efficiently we describe a
simple SG algorithm to estimate the amplitude of user $k$, as
given by
\begin{equation}
\begin{split}
{\hat{A}}_{k}[i+1] & = {\hat{A}}_{k}[i] - \mu_{A} b_k^*[i]\hat{\bf
h}^H[i] {\bf C}_k^H  \Big( {\hat{A}}_{k}[i] b_k[i] {\bf C}_k
\hat{\bf h}[i] - {\bf r}[i] \Big).
\end{split}
\end{equation}

\textit{Interference Estimation and Tracking:}\\

Let us consider the RAKE receiver with perfect channel knowledge,
whose parameter vector ${\bf f}_{k}[i]= {\bf C}_{k}{\bf h}[i]$ for
user $k$ (desired one) estimates the effective signature sequence
at the receiver, i.e. ${\bf C}_{k} {\bf h}[i]= \tilde{\bf
s}_{k}[i]$. The output of the RAKE receiver is given by:
\begin{equation}
\begin{split}
x_{k}[i]  = {\bf f}_{k}^{H}[i]{\bf r}[i] & = \underbrace{A_{k}[i]
b_{k}[i] {\bf f}_{k}^{H}[i]\tilde{\bf s}_{k}[i]}_{\rm desired
~signal} + \underbrace{\sum_{\substack{j=2 \\ j \neq k}}^{K}
A_{j}[i]b_{j}[i] {\bf f}_{j}^{H}[i]\tilde{\bf s}_{j}[i]}_{\rm MAI} \\
& \quad + \underbrace{{\bf f}_{k}^{H}[i]\boldsymbol{\eta}[i]}_{\rm
ISI} + \underbrace{{\bf f}_{k}^{H}[i]{\bf n}[i]}_{\rm noise},
\end{split}
\end{equation}
where ${\bf f}_{k}^{H}[i]\tilde{\bf s}_{k}[i]=\rho_k[i]$ and ${\bf
f}_{k}^{H}[i]\tilde{\bf s}_{j}[i]= \rho_{1,j}[i]$ for $j \neq 1$.
The symbol $\rho_k$ represents the cross-correlation (or inner
product) between the effective signature and the RAKE with perfect
channel estimates. The symbol $\rho_{1,j}[i]$ represents the
cross-correlation between the RAKE receiver and the effective
signature of user $j$. The second-order statistics of the output
of the RAKE in (31) are described by:
\begin{equation}
\begin{split}
E[|x_{k}[i]|^{2}] & = A_{k}^2[i] \rho_p^2[i]
\underbrace{E[|b_{k}[i]|^2]}_{\rightarrow 1} \\ & \quad +
\underbrace{\sum_{\substack{j=1 \\ j \neq k}}^{K}
~~\sum_{\substack{l=1 \\ j \neq k}}^{K}
A_{j}^2[i]E[b_{j}[i]b_{l}^{*}[i]] {\bf f}_{j}^{H}\tilde{\bf
s}_{j}\tilde{\bf s}_{l}^{H}{\bf f}_{j}}_{\rightarrow \sum_{j=1,j
\neq k}^{K}{\bf f}_{j}^{H}\tilde{\bf s}_{j}\tilde{\bf
s}_{j}^{H}[i]{\bf f}_{j}} \\ & \quad + {{\bf f}_{k}^{H}
E[\boldsymbol{\eta}[i]\boldsymbol{\eta}^{H}[i]]{\bf f}_{k}} +
\underbrace{{\bf f}_{k}^{H}E[{\bf n}[i]{\bf n}^{H}[i]]{\bf
f}_{k}}_{\rightarrow \sigma^2{\bf f}_{k}^{H}{\bf f}_{k}}.
\end{split}
\end{equation}

From the analysis above, we conclude that through the second-order
statistics one can identify the sum of the power levels of MAI,
ISI and the noise terms. Therefore, our strategy is to obtain
instantaneous estimates of the MAI, the ISI and the noise from the
output of a RAKE receiver, subtract the detected symbol in (3)
from this output (using the more reliable multiuser receiver
(${\bf w}_{k}[i]$)) and to track the interference (MAI + ISI +
noise) power as shown in Fig. 2. Let us define the difference
between the output of the RAKE receiver and the detected symbol
for user $1$:
\begin{equation}
\begin{split}
d_{k}[i]  = x_{k}[i] - \hat{A}_{k}[i]\hat{b}_{k}[i] & \approx
\underbrace{\sum_{k=2}^{K} A_{k}[i]b_{k}[i] {\bf
f}_{k}^{H}[i]\tilde{\bf s}_{k}[i]}_{\rm MAI} + \underbrace{{\bf
f}_{k}^{H}[i]\boldsymbol{\eta}[i]}_{\rm ISI} \\ & +
\underbrace{{\bf f}_{k}^{H}[i]{\bf n}[i]}_{\rm noise}.
\end{split}
\end{equation}
By taking expectations on $|d_{k}[i]|^2$ and taking into account the
assumption that MAI, ISI and noise are uncorrelated we have:
\begin{equation}
\begin{split}
E[|d_{k}[i]|^{2}] & \approx  \sum_{k=2}^{K}{\bf
f}_{k}^{H}[i]\tilde{\bf s}_{k}[i]\tilde{\bf s}_{k}^{H}[i]{\bf
f}_{k}[i] \\ & + {{\bf f}_{k}^{H}[i]
E[\boldsymbol{\eta}[i]\boldsymbol{\eta}^{H}[i]]}{\bf f}_{k}[i] +
\sigma^2{\bf f}_{k}^{H}[i]{\bf f}_{k}[i],
\end{split}
\end{equation}
where the above equation represents the interference power. Based on
time averages of the instantaneous values of the interference power,
we introduce the following algorithm to estimate and track
$E[|d_{k}[i]|^{2}]$:
\begin{equation}
\hat{v}[i+1] = (1-\beta) \hat{v}[i] +  \beta |d_{k}[i]|^2,
\end{equation}
where $\beta$ is a forgetting factor.  To incorporate parameter
dependency and interference power for computing a more effective
bound, we propose the parameter and interference dependent bound
(PIDB):
\begin{equation}
\gamma_k[i+1] = (1-\beta) \gamma_k[i] + \beta \Big(\sqrt{ \tau
~{\hat v}^2[i]} + \sqrt{\alpha||{\bf w}_{k}||^{2}
{\hat{\sigma}}^{2}_{v}[i]}\Big),
\end{equation}
where ${\hat v}[i]$ is the estimated interference power in the
multiuser system and $\tau$ is a weighting parameter that must be
set. Similarly to (26), the equations in (35) and (36) are
time-averaged recursions that are aimed at tracking the quantities
$|d_k[i]|^2$ and $(\sqrt{ \tau \hat{v}^2[i]} + \sqrt{\alpha ||{\bf
w}_k ||^2 \hat{\sigma_v}^2[i]} )$, respectively. The equations in
(35) and (36) also avoid undesirable too high or low instantaneous
values which may lead to inappropriate time-varying bound
$\gamma_k[i]$.

\section{Analysis of The Algorithms}

In this section we analyze the channel estimation and interference
estimation algorithms described in the previous section. We focus
on the convergence properties of the algorithms in terms of the
step size parameters $\mu_h$, $\mu_A$ and $\beta$ used for the
channel, the amplitude and the interference power estimators,
respectively.

\subsection{Channel Estimator}

In order to analyze the SG channel estimator given in (28), let us
first define an error vector $\boldsymbol{\epsilon}_h [i] =
\hat{\bf h}[i] - {\bf h}_{opt}$. By subtracting ${\bf h}_{opt} $
from the SG recursion in (28), we get
\begin{equation}
\begin{split}
\boldsymbol{\epsilon}_h [i+1] & = \boldsymbol{\epsilon}_h [i] -
\mu_h \hat{A}_k[i]{\bf C}_k^H b_k^*[i] \big( \hat{A}_k[i]b_k[i]
{\bf C}_k \hat{\bf h}[i] \\  & \quad - {\bf r}[i] \big) \\ \quad &
= \boldsymbol{\epsilon}_h [i] - \mu_h \hat{A}_k[i] b_k^*[i]{\bf
C}_k^H  \big( \hat{A}_k[i] b_k[i] {\bf C}_k (
\boldsymbol{\epsilon}_h [i] + {\bf h}_{opt}) \\  & \quad - {\bf
r}[i] \big)
\\ \quad & = \big[ {\bf I} - \mu_h \hat{A}_k[i] |b_k[i]|^2{\bf C}_k^H {\bf C}_k  \big] \boldsymbol{\epsilon}_h [i]  \\
& \quad - \mu_h \hat{A}_k[i] b_k^*[i]{\bf C}_k^H \big(
\hat{A}_k[i] b_k[i]{\bf C}_k  {\bf h}_{opt}  - {\bf r}[i] \big) \\
& = \big[ {\bf I} - \mu_h \hat{A}_k[i] |b_k[i]|^2{\bf C}_k^H {\bf
C}_k  \big] \boldsymbol{\epsilon}_h [i] \\  & \quad - \mu_h
\hat{A}_k[i] b_k^*[i]{\bf C}_k^H {\bf e}_{k,\textrm{opt}}[i].
\end{split}
\end{equation}
By considering that the error vector $\boldsymbol{\epsilon}_h
[i]$, ${\bf e}_{k,\textrm{opt}}[i]$, the signal components
$\sum_{k=1}^K {\bf x}_k[i]$ from the data vector ${\bf r}[i] =
\sum_{k=1}^K {\bf x}_k[i] + {\bf n}[i]$ given by (1) and ${\bf
n}[i]$ are statistically independent and computing the covariance
matrix of the error vector, i.e. ${\bf K}[i] = E\big[
\boldsymbol{\epsilon}_h [i] \boldsymbol{\epsilon}_h^H [i]  \big]$,
we obtain
\begin{equation}
\begin{split}
{\bf K}[i+1] & = \big[ {\bf I} - \mu_h \sigma_{A_{k}}^2\sigma_b^2
{\bf C}_k^H {\bf C}_k  \big] {\bf K}[i] \big[ {\bf I} - \mu_h
\sigma_{A_{k}}^2\sigma_b^2 {\bf C}_k^H {\bf C}_k \big] \\ & \quad
+ \mu_h^2 \sigma_{A_{k}}^2\sigma_b^2 ||{\bf C}_k^H {\bf C}_k ||^2
\textrm{MSE}_{min},
\end{split}
\end{equation}
where $\sigma_b=E[|b_k[i]|^2]$, $\sigma_{A_{k}}^2 = E[|A_{k}[i]
|^2]$ and $\textrm{MSE}_{min} = E[||{\bf
e}_{k,\textrm{opt}}[i]||^2]$ stands for the minimum MSE achieved
by the estimator. The recursive rule/algorithm in (28) is
asymptotically unbiased and will converge to the optimum channel
estimator ${\bf h}_{\textrm opt}$ if the step size $\mu_h$
satisfies the following condition
\begin{equation}
0 < \mu_h < \frac{2}{\sigma_b^2 \sigma_{A_{k}}^2 \lambda_{max}},
\end{equation}
where $\lambda_{max}$ is the largest eigenvalue of the matrix
${\bf C}_k^H {\bf C}_k$. The condition above with concern to the
step size $\mu_h$ arises from difference equations. The quantities
generated in (38) represent a geometric series with a geometric
ratio equal to $(1- \mu_h \sigma_{A_{k}}^2 \sigma_b^2 {\bf C}_k^H
{\bf C}_k)$. For stability or convergence of this algorithm, the
magnitude of this geometric ratio must be less than $1$ for all
$k$ ($| 1- \mu_h \sigma_{A_{k}}^2 \sigma_b^2 {\bf C}_k^H {\bf
C}_k|<1$). This means that $0 < \mu_h < \frac{2}{\sigma_b^2
\sigma_{A_{k}}^2 \lambda_{max}}$. The reviewer is referred to
\cite{diniz,bert} for further details.

\subsection{Amplitude Estimator}

In order to analyze the SG amplitude estimator described in (30),
let us first define an error signal $\epsilon_{A,k} [i] =
\hat{A}_k[i] - A_{k,opt}$. By subtracting $A_{k,opt}$ from the
equation in (30) we obtain
\begin{equation}
\begin{split}
\epsilon_{A_k}[i+1] & = \epsilon_{A_k}[i] - \mu_A b_k^*[i]
\hat{\bf h}^H[i] {\bf C}_k^H \big( \hat{A}_k[i] b_k[i] {\bf C}_k
\hat{\bf h}[i] - {\bf r}[i] \big)   \\
& = \epsilon_{A_k}[i] - \mu_A b_k^*[i] \hat{\bf h}^H[i] {\bf
C}_k^H \big( ( \epsilon_{A,k} [i] \\  & \quad + A_{k,opt} )
b_k[i] {\bf C}_k
\hat{\bf h}[i] - {\bf r}[i] \big)   \\
& = \epsilon_{A_k}[i] - \mu_A |b_k[i]|^2 \hat{\bf h}[i] {\bf
C}_k^H {\bf C}_k \hat{\bf h}[i]\epsilon_{A_k}[i] \\ & \quad -
\mu_A b_k^*[i] \hat{\bf h}^H[i] {\bf C}_k^H \big(  A_{k,opt}
b_k[i] {\bf C}_k \hat{\bf h}[i] - {\bf r}[i] \big)  \\ & = \big(1
- \mu_A |b_k[i]|^2  \hat{\bf h}[i] {\bf C}_k^H {\bf C}_k \hat{\bf
h}[i] \big)\epsilon_{A_k}[i] \\  & \quad - \mu_A b_k^*[i] \hat{\bf
h}^H[i] {\bf C}_k^H e_{A,\textrm{opt}}[i].
\end{split}
\end{equation}
By considering that the error $\epsilon_A[i]$, ${\bf
e}_{A,\textrm{opt}}[i]=A_{k,opt} b_k[i] {\bf C}_k \hat{\bf h}[i] -
{\bf r}[i]$, the signal components $\sum_{k=1}^K {\bf x}_k[i]$
from the data vector ${\bf r}[i] = \sum_{k=1}^K {\bf x}_k[i] +
{\bf n}[i]$ given by (1) and ${\bf n}[i]$ are statistically
independent, and by computing the mean-squared error, i.e.,
$\textrm{MSE}(A_k[i] ) [i] = K_{A_k} [i] = E[ |\epsilon[i]|^2]$,
we get
\begin{equation}
\begin{split}
\textrm{MSE}(A_k[i] )[i+1]  & = \big(1 - \mu_A \sigma_b^2 E[||{\bf
h}_k^H[i] {\bf C}_k^H ||^2] \big)^2 K_{A_k}[i]
\\ &  \quad + \mu_A^2 \sigma_b^2 E[||{\bf h}_k^H[i] {\bf C}_k^H
||^2] \textrm{MSE}_{A,min},
\end{split}
\end{equation}
where $\textrm{MSE}_{A,min} = E[||{ e}_{A,\textrm{opt}}[i]||^2]$.
The cross multiplication between the terms will vanish as a result
of the statistical independence between them. The general
algorithm in (30) will converge asymptotically and in an unbiased
way to the $A_{k,opt}$ provided the step size $\mu_A$ is chosen
such that
\begin{equation}
0 < \mu_A < \frac{2}{ \sigma_b^2 E[||{\bf h}_k^H[i] {\bf C}_k^H
||^2]}.
\end{equation}
The above range of values has to be tuned in order to ensure good
convergence and tracking of the amplitude.

\subsection{Interference Estimation and Tracking
Algorithm}

Let us describe in a general form the mechanisms for interference
estimation and tracking given in (26) and (36). We can write
without loss of generality
\begin{equation}
\gamma_k[i+1] = (1 - \beta) \gamma_k[i] + \beta ~\textrm{Po}[i],
\end{equation}
where $\textrm{Po}[i]$ can account either for a parameter
dependent bound (PDB), as described in Section IV. A, or for a
parameter and interference dependent bound (PIDB), as the one
detailed in Section IV.B.

Our goal is to establish the convergence of a general stochastic
recursion, as the one given in (43), in terms of the mean-squared
error (MSE) at iteration $[i]$, as described by
\begin{equation}
\textrm{MSE}(\gamma[i])[i] = E \big[ | \gamma_k [i] -
\gamma_{k,\textrm{opt}} |^2 \big],
\end{equation}
where $\gamma_{k,\textrm{opt}}$ is the optimal parameter estimate
for $\gamma_k[i]$.

We will show that under certain conditions on $\beta$, the
sequence converges to the optimal $\gamma_{k,\textrm{opt}}$ in the
mean-square sense. Let us define the error
\begin{equation}
\epsilon_{\gamma} [i] = \gamma_k[i] - \gamma_{k,\textrm{opt}},
\nonumber
\end{equation}
and substitute the above equation into (44), which yields
\begin{equation}
\begin{split}
\epsilon_{\gamma} [i+1] & = (1- \beta) \epsilon_{\gamma} [i] +
\beta (\textrm{Po}[i] -  \gamma_{k,\textrm{opt}}) \\
& = (1- \beta) \epsilon_{\gamma} [i] + \beta e_{k,\textrm{opt}}
[i],
\end{split}
\end{equation}
The MSE at time instant $[i+1]$ is given by
\begin{equation}
\begin{split}
\textrm{MSE}(\gamma[i+1])[i+1] & = \epsilon_{\gamma}^*[i+1]
\epsilon_{\gamma}[i+1] = | \epsilon_{\gamma} [i+1] |^2 \\ & = (1-
\beta)^2 \epsilon_{\gamma}^*[i] \epsilon_{\gamma}[i] \\ & \quad +
\beta^2 ~e_{k,\textrm{opt}}[i]e_{k,\textrm{opt}}^*[i] \\ & \quad +
(1-\beta) \beta ~\epsilon_{\gamma}[i]  e_{k,\textrm{opt}}^*[i] \\
& \quad + (1- \beta) \beta ~\epsilon_{\gamma}^*[i]
e_{k,\textrm{opt}}[i].
\end{split}
\end{equation}
By taking the expected value on both sides and assuming that
$\epsilon_{\gamma} [i]$ and $e_{k,\textrm{opt}}[i]$ are
statistically independent we have
\begin{equation}
\begin{split}
E \big[\textrm{MSE}[i+1] \big] & = E \big[| \epsilon_{\gamma}
[i+1] |^2 \big] \\ & =  K_{\gamma}[i+1] \\ & = (1- \beta)
K_{\gamma} [i] (1- \beta) + \beta^2 |e_{k,\textrm{opt}}[i]|^2,
\end{split}
\end{equation}
where $ |e_{k,\textrm{opt}}[i]|^2$ is the minimum MSE achieved by
the estimator.

We can notice that the cross-multiplication terms will vanish as a
result of the statistical independence between the terms. The
general recursive rule in (43) will converge asymptotically
provided the step size $\beta$ is chosen such that
\begin{equation}
0 < \beta < 2.
\end{equation}
The above range of values has to be adjusted in order to ensure
good convergence and tracking of the parameter dependency and/or
the interference modeled here as the quantity $\textrm{Po}[i]$.

\section{Simulations}

In this section we assess the performance of the proposed and
existing adaptive algorithms in several scenarios of practical
interest:
\begin{itemize}
\item{The NLMS, the SM-NLMS \cite{huang}, the SM-NLMS of Guo and Huang
\cite{guo} and the proposed SM-NLMS with the parameter-dependent
(PDB) and parameter and interference dependent (PIDB) time-varying
bounds.}

\item{The AP \cite{diniz}, the SM-AP \cite{werner1} and the proposed
SM-AP with the PDB and PIDB time-varying bounds.}

\item{The BEACON \cite{nagaraj} and the proposed BEACON algorithms
with the PDB and PIDB time-varying bounds.}

\end{itemize}

We consider for the sake of simplicity binary phase-shift-keying
(BPSK) modulation, a DS-CDMA system with Gold sequences of length
$N=31$ and typical fading channels found in mobile communications
systems that can be modeled according to Clarke's model
\cite{rappa}. The channels experienced by different users are
identical since we focus on a downlink scenario and the desired
receiver processes the transmissions intended to other users (MAI)
over the same channel as its own signal. The channel coefficients
are $h_{l}[i]=p_{l}\alpha_{l}[i]$, where $\alpha_{l}[i]$
($l=0,1,\ldots,L_{p}-1$) are obtained with Clarke's model
\cite{rappa}. In particular, we employ standard SG adaptive
algorithms for channel estimation and RAKE design in order to
concentrate on the comparison between the analyzed algorithms for
receiver parameter estimation. We show the results in terms of the
normalized Doppler frequency $f_{d}T$ (cycles/symbol) and use
three-path channels with relative powers given by $0$, $-3$ and
$-6$ dB, where in each run the spacing between paths is obtained
from a discrete uniform random variable between $1$ and $2$ chips.
The channel coefficients $h_{l}[i]$ ($l=0,1,\ldots,L_{p}-1$) are
constant during each symbol interval and change according to
Clarke's model over time. Since the channel is modelled as an FIR
filter, we employ a channel estimation filter with $6$ taps as an
upper bound for the experiments. Note that the delays of the
channel taps are multiples of the chip rate and are random. Their
range coincides with the maximum length of the estimation filter
which is $6$ taps.

The parameters of the algorithms are optimized  and shown for each
example, the system has a power distribution for the interferers
that follows a log-normal distribution with associated standard
deviation of $3$ dB and experiments are averaged over $100$
independent runs. The receivers are trained with $200$ symbols and
then switch to decision-directed mode for each data packet. We
address the dynamic channel by adjusting the receiver weights with
the training sequences (with length equal to $200$ symbols) and
then we exploit the decision-directed mode to track the channel
variations. If the channel does not vary too fast then the
adaptive receivers can track it with this scheme, as will be shown
in what follows.

\subsection{Interference Estimation and Tracking}

In order to evaluate the effectiveness of the proposed
interference estimation and tracking algorithms, that are
incorporated into the time-varying error bounds for tracking the
MAI and ISI powers, we carried out an experiment, depicted in Fig.
2. In this scenario, the proposed algorithm estimates of the MAI
and ISI powers are compared to the actual interference power
levels that are generated by the simulations. The results obtained
show that the proposed algorithms are very effective for
estimating and tracking the interference power in dynamic
environments. Specifically, the estimation error does not exceed
in average $5 \%$  of the estimated power level, as depicted in
Fig. 2. Indeed, the algorithms are capable of accurately
estimating the interference power levels and tracking them, which
can be verified by the proximity between the curves obtained with
the estimation algorithms and the actual values.

\begin{figure}[!htb]
\begin{center}
\def\epsfsize#1#2{1\columnwidth}
\epsfbox{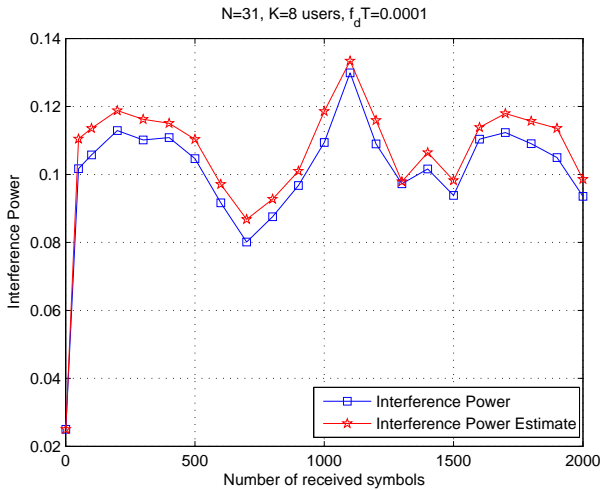} \caption{Performance of the interference power
estimation and tracking at $E_{b}/N_{0}=12$ dB and with $\beta =
0.05$ .}
\end{center}
\end{figure}

\subsection{SINR Performance}

In this part, the performance of the proposed algorithms is
assessed in terms of output signal-to-interference-plus-noise
ratio (SINR), which is defined as
\begin {equation}
\centering {\rm SINR}[i]=\frac{{\boldsymbol w}^{H}[i]{\boldsymbol
R}_{s}[i]{\boldsymbol w}[i]}{{\boldsymbol w}^{H}[i]{\boldsymbol
R}_{I}[i]{\boldsymbol w}[i]},
\end{equation}
where ${\boldsymbol R}_{s}[i]$ is the autocorrelation matrix of
the desired signal and ${\boldsymbol R}_{I}[i]$ is the
cross-correlation matrix of the interference and noise in the
environment. The goal here is to evaluate the convergence
performance of the proposed algorithms with time-varying error
bounds for the modified SM-NLMS, SM-AP and BEACON techniques.
Specifically, we consider examples where the adaptive receivers
converge to about the same level of SINR, which illustrates in a
fair way the speed of convergence of the proposed algorithms and
the existing ones. We also measure the update rate (UR) of all the
SM-based algorithms as an important complexity issue.

The SINR convergence performance of NLMS, AP and BEACON algorithms
is illustrated via computer experiments in Figs. 3, 4 and 5,
respectively. The curves in Fig. 3 show that the proposed SM-NLMS
algorithms with the PIDB time-varying error bounds achieve the
fastest convergence, followed by the proposed SM-NLMS-PDB, the
SM-NLMS-Huang \cite{guo}, the conventional SM-NLMS \cite{huang} and
the NLMS \cite{diniz} recursions. Even though the proposed
SM-NLMS-PIDB and SM-NLMS-PDB algorithms enjoy the fastest
convergence rates, they exhibit remarkably lower UR properties,
saving significant computational resources and being substantially
more economical than the conventional SM-NLMS algorithm.

\begin{figure}[!htb]
\begin{center}
\def\epsfsize#1#2{1\columnwidth}
\epsfbox{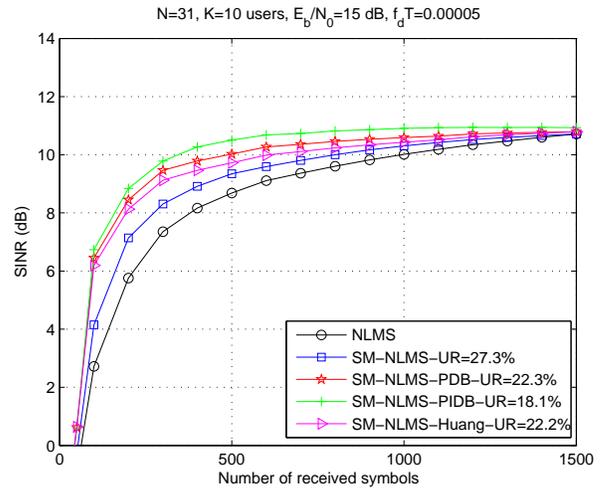} \caption{SINR Performance of NLMS algorithms at
$E_{b}/N_{0}=15$ dB with $\beta=0.05$, $\alpha=8$ and $\tau = 2$ .}
\end{center}
\end{figure}

\begin{figure}[!htb]
\begin{center}
\def\epsfsize#1#2{1\columnwidth}
\epsfbox{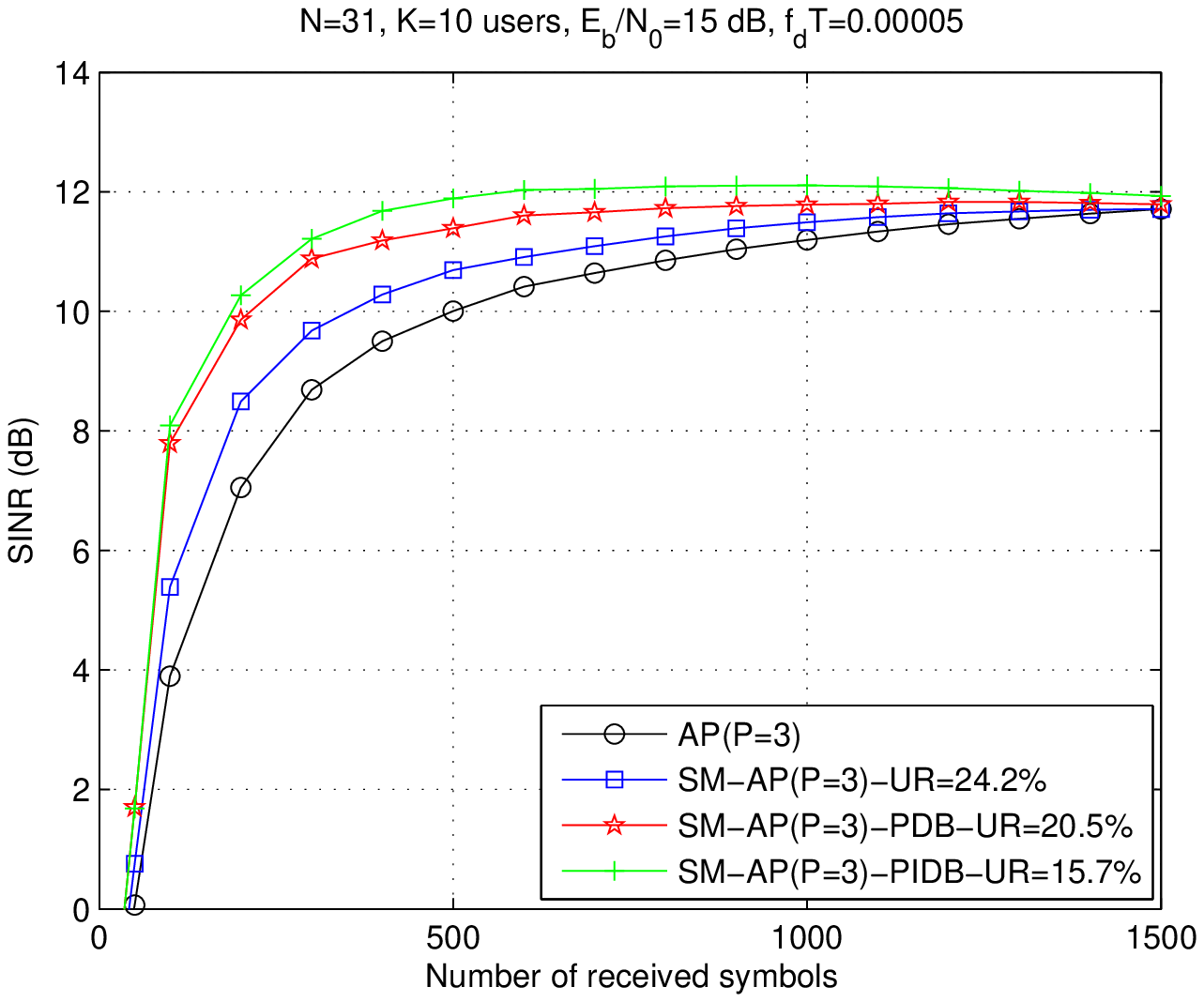} \caption{SINR Performance of AP algorithms at
$E_{b}/N_{0}=15$ dB with $\beta=0.05$, $\alpha=8$ and $\tau = 2$ .}
\end{center}
\end{figure}

\begin{figure}[!htb]
\begin{center}
\def\epsfsize#1#2{1\columnwidth}
\epsfbox{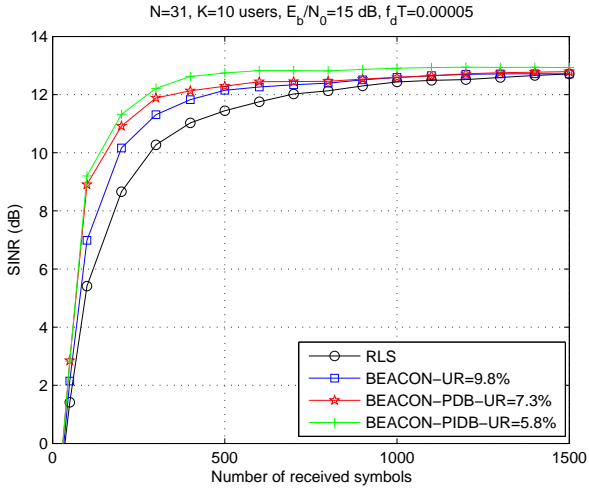} \caption{SINR Performance of RLS and BEACON
algorithms at $E_{b}/N_{0}=15$ dB with $\beta=0.05$, $\alpha=5$ and
$\tau = 1.5$ .}
\end{center}
\end{figure}

By observing the results for the AP and the BEACON algorithms,
shown in Figs. 4 and 5, one can notice that the results
corroborate those found for the NLMS algorithms. It should be
noted that despite their higher complexity than NLMS algorithms,
the AP and BEACON techniques have faster convergence, better SINR
steady-state performance and lower URs.

\subsection{BER Performance}

In this subsection, we focus on the bit error rate (BER)
performance of the proposed algorithms. We consider a simulation
setup where the data packets transmitted have $1500$ symbols and
the adaptive receivers and algorithms are trained with $200$
symbols and then switch to decision-directed mode, in which they
continue to adapt and track the channel variations.

\begin{figure}[!htb]
\begin{center}
\def\epsfsize#1#2{1\columnwidth}
\epsfbox{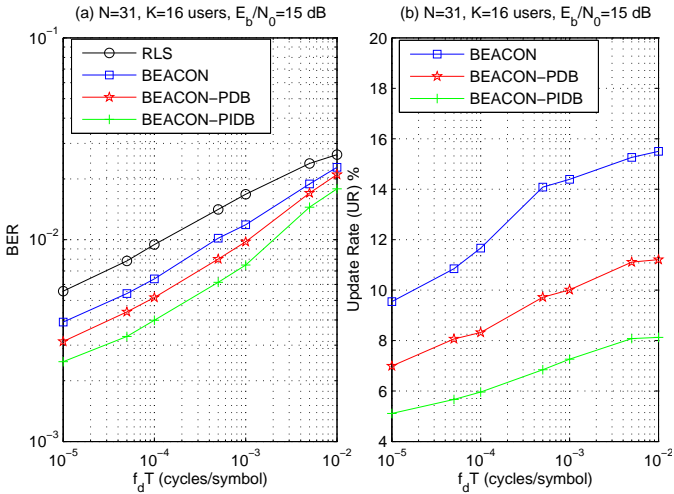} \caption{(a) BER performance versus $f_dT$ and
(b) Update Rate (UR) versus $f_dT$ for the BEACON algorithms for a
forgetting factor $\lambda=0.997$ for the conventional RLS
algorithm. The proposed BEACON-PDB abd BEACON-PIDB algorithms use
$\beta=0.05$, $\alpha=5$ and $\tau = 1.5$.}
\end{center}
\end{figure}

Firstly, we consider a study of the BER performance and the impact
on the UR of the fading rate of the channel ($f_dT$) in the
experiment shown in Fig. 6. We observe from the curves in Fig. 6 (a)
that the new BEACON algorithms obtain substantial gains in BER
performance over the original BEACON \cite{nagaraj} and the RLS
algorithm \cite{diniz} for a wide range of fading rates. In
addition, as the channel becomes more hostile the performance of the
analyzed algorithms approaches one another, indicating that the
adaptive techniques are encountering difficulties in dealing with
the changing environment and interference. With regard to the UR,
the curves in Fig. 6 (b) illustrate the impact of the fading rate on
the UR of the algorithms. Indeed, it is again verified that the
proposed BEACON-PDB and BEACON-PIDB algorithms can obtain
significant savings in terms of UR, allowing the mobile receiver to
share its processing power with other important functions and to
save battery life.

\begin{figure}[!htb]
\begin{center}
\def\epsfsize#1#2{1\columnwidth}
\epsfbox{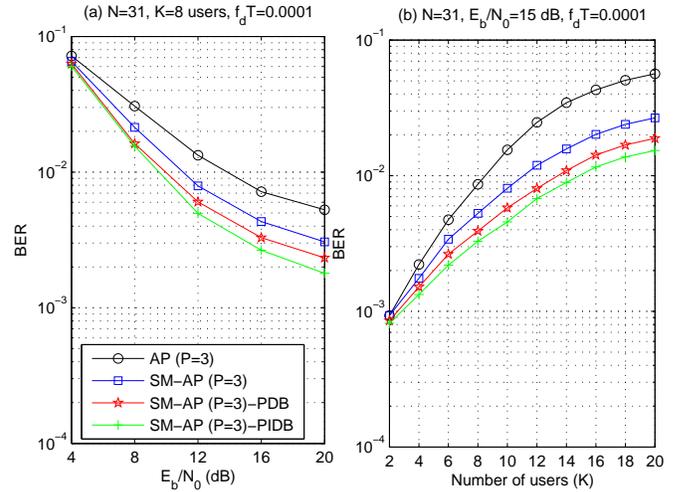} \caption{BER performance versus (a) $E_{b}/N_{0}$
dB (b) $K$ for AP algorithms with $P=3$.}
\end{center}
\end{figure}

The BER performance versus the signal-to-noise ratio
($E_{b}/N_{0}$) and the number of users (K) is illustrated in
Figs. 7 for the  AP with $P=3$. The results confirm the excellent
performance of the proposed PIDB time-varying error bound for a
variety of scenarios, algorithms and loads. The PIDB technique
allows significantly superior performance while reducing the UR of
the algorithm and saving computations. We can also notice that
significant performance and capacity gains can be obtained by
exploiting data reuse. From the curves it can be seen that the
proposed PIDB mechanism with the SM-AP with $P=3$ can save up to
$3$ dB and up to $4$ dB as compared to the PDB and to the SM-AP
with fixed bounds, respectively, for the same BER performance. In
terms of system capacity, we verify that the PIDB approach can
accommodate up to $4$ more users as compared to the PDB technique
for the same BER performance.

\section{Conclusions}

We proposed SM adaptive algorithms based on time-varying error
bounds. Adaptive algorithms for tracking MAI and ISI power and
taking into account parameter dependency were incorporated into
the new time-varying error bounds. Simulations show that the new
algorithms outperform previously reported techniques and exhibit a
reduced number of updates. The proposed algorithms can have a
significant impact on the design of low-complexity receivers for
spread spectrum systems, as well as for future MIMO systems
employing either CDMA or OFDM as the multiple access technology.
The proposed algorithms are especially relevant to future wireless
cellular, ad hoc and sensor networks, where their potential to
save computational resources may play a significant role given the
limited battery resources and processing capabilities of mobile
units and sensors.


\end{document}